\documentclass{aa}

\usepackage{epsf}

\begin{document}
   \thesaurus{08         
              (01.01.2;
		02.02.1;  
               13.07.2;  
               08.02.3)} 

\title{{\em RXTE} observations of XTE~J2012+381 during its 1998 outburst.}

\author{L.~Vasiliev\inst{1}, S.~Trudolyubov\inst{2,1}, 
M.~Revnivtsev\inst{1,3}}

\offprints{leonid@hea.iki.rssi.ru}

\institute{Space Research Institute, Russian Academy of Sciences
Profsoyuznaya 84/32, 117810 Moscow, Russia
\and Los Alamos National Laboratory, Los Alamos, 87545 New Mexico, USA
\and Max-Planck-Institut f\"ur Astrophysik,
Karl-Schwarzschild-Str. 1, 85740 Garching bei Munchen, Germany}

   \date{}

\maketitle
\markboth{Vasiliev, Trudolyubov \& Revnivtsev}{{\em RXTE} observations of 
XTE~J2012+381 during its 1998 outburst.} 

\begin{abstract}
We present an analysis of the {\em RXTE} observations of X-ray transient 
source XTE~J2012+381 during its 1998 outburst. The spectral and timing 
properties of the source emission and their evolution during the outburst 
are very similar to what of the X-ray Novae. 

\keywords{stars: binaries: general -- stars: individual: XTE~J2012+381 --
X-rays: stars} 
\end{abstract}

\section{Introduction}
The XTE~J2012+381 was discovered as a new transient source on May 27, 1998
by All Sky Monitor (ASM) aboard the Rossi X-ray Timing Explorer, {\em RXTE} 
(Bradt, Swank $\&$ Rothschild 1993) observatory (Remillard et al. 1998). The 
X-ray source was localized with an accuracy of 1 arcmin in multiple scans of 
the region by PCA/{\em RXTE} experiment (Marshall $\&$ Strohmayer 1998): $R.A. = 20^{h}12^{m}43^{s}, Dec = 38^{\circ} 11.0 \arcmin$ (equinox 2000.0). 
Observations in the optical band by 1.0-m Jacobus Kapteyn Telescope showed 
the presence of possible optical counterpart USNO 1275.13846761 : $R.A. = 
20^{h}12^{m} 37.8^{s}, Dec = +38^{\circ} 11 \arcmin 00.6 \arcsec$ (equinox 
2000.0)(Hynes \& Roche 1998). Observations in the radio band by VLA revealed 
the presence of the likely counterpart of XTE~J2012+381: $R.A. = 20^{h} 
12^{m} 37.67, Dec = +38^{\circ}11 \arcmin 01.2\arcsec$ (equinox 2000.0, 
uncertainty $0.5\arcsec$) as well. Significant variability of the new radio 
source supports its association with X-ray transient (Hjellming et al. 1998).

Pointed instruments of {\em RXTE} -- PCA and HEXTE observed the source quasi 
evenly from May to July 1998, providing a good coverage of the whole outburst. The flux from the source was above ASM/{\em RXTE} detection limit until 
December 2, 1998. In this Letter we present the results of a spectral and 
timing analysis of {\em RXTE} pointed observations.

\begin{table}
\small
\caption{The list of {\it RXTE}/PCA observations of XTE~J2012+381 used for 
the analysis. 
\label{obslog}} 
\tabcolsep=0.1cm
\begin{tabular}{ccccc}
\hline
Obs.ID & \multicolumn{3}{c}{Time start}& Exp.$^a$, s\\
       & Date&  Time, UT &  TJD\\
\hline
 30188-04-01-04S$^{b}$&  27/05/98 & 16:21 &10960.7&   78\\
 30188-04-02-00 &  29/05/98 & 17:48 &10962.7& 3195\\
 30188-04-03-00 &  30/05/98 & 19:25 &10963.8& 6002\\
 30188-04-04-00 &  31/05/98 & 21:41 &10964.9& 2712\\
 30188-04-05-00 &  01/06/98 & 17:47 &10965.7& 6002\\
 30188-04-06-00 &  02/06/98 & 21:41 &10966.9&  780\\
 30188-04-07-00 &  03/06/98 & 18:28 &10967.8&  779\\
 30188-04-08-00 &  04/06/98 & 21:50 &10968.9&  219\\
 30188-04-09-00 &  05/06/98 & 22:43 &10970.0&  843\\
 30188-04-10-00 &  07/06/98 & 01:52 &10971.1& 2904\\
 30188-04-11-00 &  07/06/98 & 19:25 &10971.8& 3017\\
 30188-04-12-00 &  08/06/98 & 19:24 &10972.8& 3156\\
 30188-04-13-00 &  09/06/98 & 21:03 &10973.8& 2940\\
 30188-04-14-00 &  11/06/98 & 00:18 &10975.0& 2765\\
 30188-04-15-00 &  15/06/98 & 00:24 &10979.0& 2400\\
 30188-04-16-00 &  16/06/98 & 17:46 &10980.7& 1532\\
 30188-04-17-00 &  22/06/98 & 21:29 &10986.9& 1527\\  
 30188-04-18-00 &  30/06/98 & 21:29 &10994.9& 865\\
 30188-04-19-00 &  30/06/98 & 23:12 &10994.9& 699\\  
 30188-04-20-00 &  06/07/98 & 18:04 &11000.7& 1300\\ 
 30188-04-21-00 &  13/07/98 & 00:52 &11007.0& 1625\\  
 30188-04-22-00 &  19/07/98 & 18:07 &11013.8& 2783\\ 
 30188-04-23-00 &  21/07/98 & 21:26 &11015.9& 2317 \\
 30188-04-24-00 &  29/07/98 & 18:40 &11023.8&  736 \\
\hline
\end{tabular}
\begin{list}{}{}
\item[$^a$] -- Deadtime corrected value of the PCA exposure
\item[$^b$] -- Slew part of observation
\end{list}
\end{table}

\begin{figure}
\epsfxsize=8cm
\epsffile[45 197 566 714]{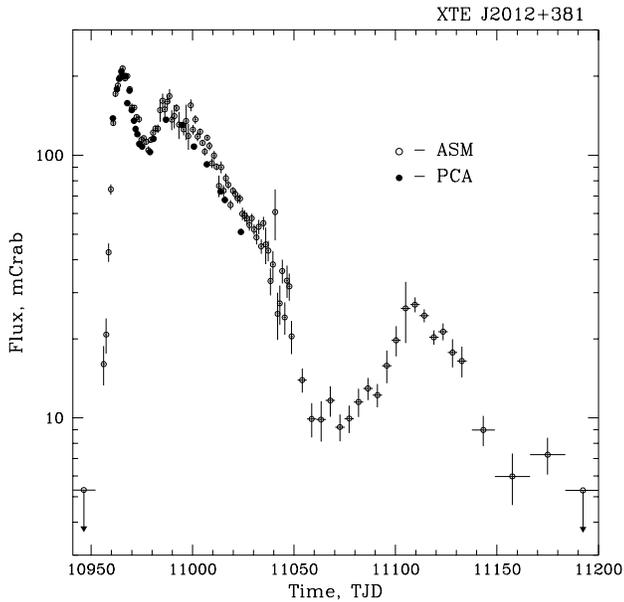}
\caption{Light curve of the 1998 outburst of X-ray transient XTE~J2012+381.
Open circles and filled circles represent the data of All Sky Monitor
(ASM/{\em RXTE}, $1.3 - 12.2$ keV) and Proportional Counter Array (PCA/{\em RXTE}, $3 - 20$ keV) respectively. \label{lcurve}} 
\end{figure}

\section{Observations and data analysis}

In our analysis we used all the publicly available data obtained from
{\em RXTE} archive including 24 pointed observations. The 24 observations 
quasi 
evenly cover the 1998 outburst of the source with a total exposure of $\sim 
46$ ks. The list of observations is presented in Table \ref{obslog}. 

For data reduction we used the standard FTOOLS package version 5.0
For the spectral analysis we used PCA data collected in the 3-20 keV
energy range. The response matrix was constructed for every observation 
using pcarmf v3.5. For the PCA background subtraction we 
applied a Very Large Events (VLE)-based model. The standard 
dead time correction procedure was applied to the PCA data. In order to 
account for the uncertainties of the response matrix, a 1$\%$ systematic 
error was added to the statistical error for each PCA energy channel.

HEXTE data reduction was done with FTOOLS 5.0 standard tasks and according
to RXTE GOF recommendations.

We generated the energy spectra of XTE~J2012+381 averaging the data over 
the whole single observations. The spectral data were approximated with the
simplest two-component model: the sum 
of the XSPEC ``multicolor disk black body'' (Mitsuda et al. 1984) and a 
simple power law model with inclusion of a broad gaussian emission line 
at 6.4 keV. The results of the spectral approximation of the PCA 
data with analytical models described above are presented in Table 
\ref{spec_pars}.

For the timing analysis of XTE~J2012+381 the PCA {\em Generic Event} 
timing mode data were used. We generated power density spectra (PDS) 
in the $0.001 - 4096$ Hz frequency range using combined data of several 
consequitive observations in order to improve statistical significance 
of the results. The resulting spectra were logarithmically rebinned to 
reduce scatter at high frequencies and normalized to square root of 
fractional variability rms. The white noise due to the Poissonian statistics 
corrected for the dead-time effects, was subtracted (\cite{vihl94}; 
\cite{zhang95}; \cite{mikej00}).

\begin{table*}
\small
\caption{Spectral parameters of XTE~J2012+381, derived using a combination 
of a multicolor disc blackbody, power law and gaussian emission line. 
Parameter errors correspond to a $1 \sigma$ confidence level for the assumed 
$1 \%$ systematic uncertainty of data. \label{spec_pars}}
\tabcolsep=0.45cm
\begin{tabular}{cccccccc}
\hline
\hline
\#& $T_{\rm in}$, keV& $R_{\rm in}\cos i$& $\alpha$&Line Width& Eqw. Width&Flux$^{b}$ 
&$\chi^2$ (dof)\\
& &km$^{a}$& & keV & eV& & \\
\hline
 1 &$0.79 \pm 0.01 $&$ 27.5 \pm  1.4 $&$ 2.2 \pm  0.1 $&$  1.4 \pm  0.1$ &$ 825 \pm  164$&$ 36.8 \pm  1.1 $& 35.1(33) \\ 
 2 &$0.79 \pm 0.01 $&$ 38.7 \pm  0.7 $&$ 2.0 \pm  0.0 $&$  1.1 \pm  0.1$ &$ 312 \pm  32 $&$ 49.6 \pm  1.5 $& 30.4(33) \\ 
 3 &$0.80 \pm 0.01 $&$ 39.6 \pm  0.8 $&$ 2.0 \pm  0.1 $&$  1.1 \pm  0.1$ &$ 297 \pm  38 $&$ 54.1 \pm  1.6 $& 38.8(33) \\ 
 4 &$0.81 \pm 0.01 $&$ 39.5 \pm  0.7 $&$ 2.0 \pm  0.0 $&$  1.2 \pm  0.1$ &$ 237 \pm  32 $&$ 56.9 \pm  1.7 $& 38.5(33) \\ 
 5 &$0.81 \pm 0.01 $&$ 39.4 \pm  0.7 $&$ 2.0 \pm  0.0 $&$  1.1 \pm  0.1$ &$ 178 \pm  28 $&$ 54.7 \pm  1.6 $& 35.8(33) \\ 
 6 &$0.80 \pm 0.01 $&$ 40.1 \pm  0.8 $&$ 2.1 \pm  0.1 $&$  1.0 \pm  0.1$ &$ 253 \pm  42 $&$ 55.0 \pm  1.6 $& 29.1(33) \\ 
 7 &$0.80 \pm 0.01 $&$ 39.1 \pm  0.8 $&$ 2.0 \pm  0.1 $&$  1.1 \pm  0.1$ &$ 302 \pm  44 $&$ 52.6 \pm  1.6 $& 24.9(33) \\ 
 8 &$0.77 \pm 0.01 $&$ 41.3 \pm  1.0 $&$ 2.3 \pm  0.1 $&$  0.8 \pm  0.2$ &$ 260 \pm  70 $&$ 49.3 \pm  1.5 $& 34.1(33) \\ 
 9 &$0.76 \pm 0.01 $&$ 40.7 \pm  0.8 $&$ 2.7 \pm  0.2 $&$  0.4 \pm  0.1$ &$ 133 \pm  31 $&$ 40.6 \pm  1.2 $& 30.1(33) \\ 
10 &$0.76 \pm 0.01 $&$ 40.0 \pm  0.7 $&$ 2.1 \pm  0.1 $&$  1.0 \pm  0.1$ &$ 229 \pm  34 $&$ 38.2 \pm  1.1 $& 20.7(33) \\ 
11 &$0.75 \pm 0.01 $&$ 40.3 \pm  0.7 $&$ 2.7 \pm  0.1 $&$  0.6 \pm  0.1$ &$ 153 \pm  29 $&$ 35.9 \pm  1.1 $& 29.0(33) \\ 
12 &$0.75 \pm 0.01 $&$ 39.2 \pm  0.8 $&$ 2.1 \pm  0.2 $&$  1.0 \pm  0.2$ &$ 314 \pm  69 $&$ 34.4 \pm  1.0 $& 33.4(33) \\ 
13 &$0.73 \pm 0.01 $&$ 40.0 \pm  0.7 $&$ 2.3 \pm  0.1 $&$  0.8 \pm  0.1$ &$ 259 \pm  37 $&$ 31.7 \pm  1.0 $& 40.9(33) \\ 
14 &$0.73 \pm 0.01 $&$ 39.1 \pm  0.7 $&$ 2.0 \pm  0.1 $&$  1.1 \pm  0.1$ &$ 406 \pm  44 $&$ 31.0 \pm  0.9 $& 32.6(33) \\ 
15 &$0.72 \pm 0.01 $&$ 39.3 \pm  0.8 $&$ 2.1 \pm  0.1 $&$  1.0 \pm  0.1$ &$ 363 \pm  48 $&$ 29.8 \pm  0.9 $& 31.9(33) \\ 
16 &$0.75 \pm 0.01 $&$ 38.3 \pm  0.7 $&$ 1.8 \pm  0.1 $&$  1.4 \pm  0.1$ &$ 469 \pm  67 $&$ 32.7 \pm  1.0 $& 48.5(33) \\ 
17 &$0.76 \pm 0.01 $&$ 40.8 \pm  0.7 $&$ 2.8 \pm  0.4 $&$  0.7 \pm  0.2$ &$ 165 \pm  44 $&$ 39.4 \pm  1.2 $& 28.2(33) \\ 
18 &$0.76 \pm 0.01 $&$ 38.6 \pm  1.1 $&$ 4.0 \pm  0.5 $&$  0.6 \pm  0.2$ &$ 117 \pm  45 $&$ 37.6 \pm  1.1 $& 30.6(33) \\ 
19 &$0.76 \pm 0.01 $&$ 39.2 \pm  0.7 $&$ 3.1 \pm  0.5 $&$  0.7 \pm  0.2$ &$ 192 \pm  50 $&$ 37.4 \pm  1.1 $& 23.7(33) \\ 
20 &$0.73 \pm 0.01 $&$ 40.7 \pm  0.7 $&$ 3.4 \pm  0.4 $&$  0.5 \pm  0.2$ &$ 136 \pm  44 $&$ 31.8 \pm  1.0 $& 22.0(33) \\ 
21 &$0.72 \pm 0.01 $&$ 39.1 \pm  0.7 $&$ 2.0 \pm  0.3 $&$  0.7 \pm  0.2$ &$ 285 \pm  61 $&$ 26.6 \pm  0.8 $& 28.8(33) \\ 
22 &$0.69 \pm 0.01 $&$ 40.0 \pm  0.9 $&$ 1.6 \pm  0.4 $&$  1.1 \pm  0.3$ &$ 469 \pm 160 $&$ 21.9 \pm  0.7 $& 23.3(33) \\ 
23 &$0.68 \pm 0.01 $&$ 39.8 \pm  0.7 $&$ 3.1 \pm  0.3 $&$  0.6 \pm  0.1$ &$ 248 \pm  58 $&$ 20.4 \pm  0.6 $& 26.2(33) \\ 
24 &$0.65 \pm 0.01 $&$ 35.3 \pm  2.6 $&$ 4.9 \pm  0.4 $&$  0.0 \pm  1.9$ &$ 114 \pm  51 $&$ 16.8 \pm  0.5 $& 25.2(33) \\ 
\hline
\end{tabular}
\begin{list}{}{}
\item[$^a$] -- assuming the source distance of 10 kpc
\item[$^b$] -- total X-ray flux in the $3 - 20$ keV energy range in units 
of $\times 10^{-10}$ erg s$^{-1}$ cm$^{-2}$
\end{list}
\end{table*}

\section{Results}

\begin{figure}
\epsfxsize=8.0cm
\epsffile[40 150 566 714]{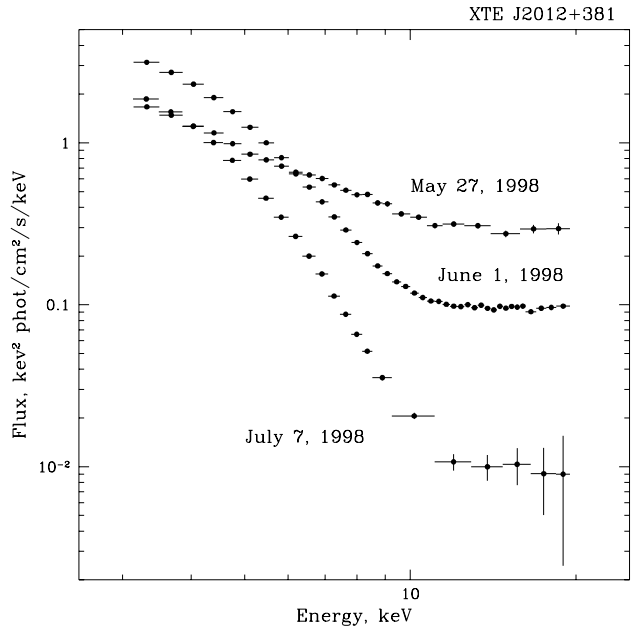}
\caption{Typical energy spectra of XTE~J2012+381 during 3 consequent
phases. One can see that the strength of the soft component rises while 
the strength of the power law component decreases. \label{3spectra}}
\end{figure}

\begin{figure}
\epsfxsize=8.0cm
\epsffile[40 197 566 714]{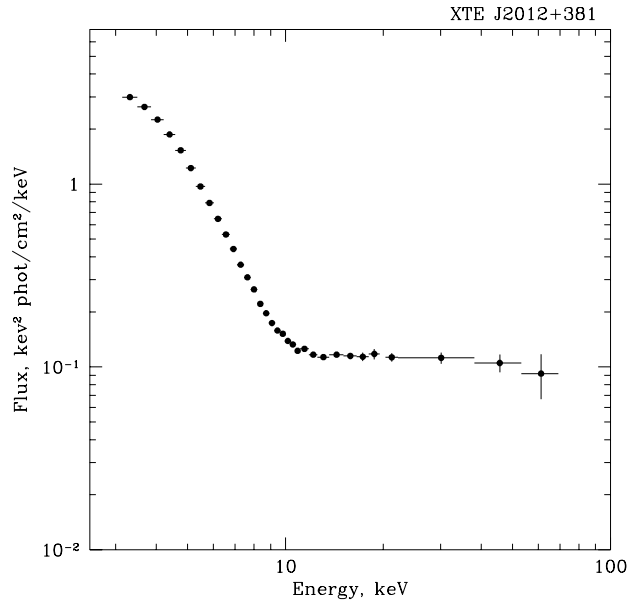}
\caption{Averaged broad band energy spectrum of XTE~J2012+381 during 
observations $\#2 - \#17$ according to the data of PCA ($3 - 20$ keV) and 
HEXTE ($20 - 80$ keV) instruments. \label{broad}}
\end{figure}

The X-ray flux history of the XTE~J2012+381 outburst based on {\em RXTE}/PCA 
and ASM data in 1.3-12 keV energy range is shown in Fig.\ref{lcurve}. The
evolution of the source flux in the soft X-ray band ($1.3 - 12$ keV as well 
as in $3 - 20$ keV energy band) was characterized by the fast initial rise 
to a level of $\sim 220$ mCrab \footnote{Assuming a distance of 10 kpc the 
luminosity in the peak of the outburst is $\sim 7 \times 10^{37}$ erg 
s$^{-1}$ in the $3 - 20$ keV energy band.} on a time scale of $\sim$ a week 
followed by a $\sim$3 day long maximum and relatively slow decay, interrupted 
by secondary maximum $\sim$30 days after the beginning of the outburst. The 
subsequent outburst evolution was unusual because of long and powerful 
tertiary peak (close to 150 days from the beginning of the outburst). 

\begin{figure}
\epsfxsize=8cm
\epsffile[40 197 566 614]{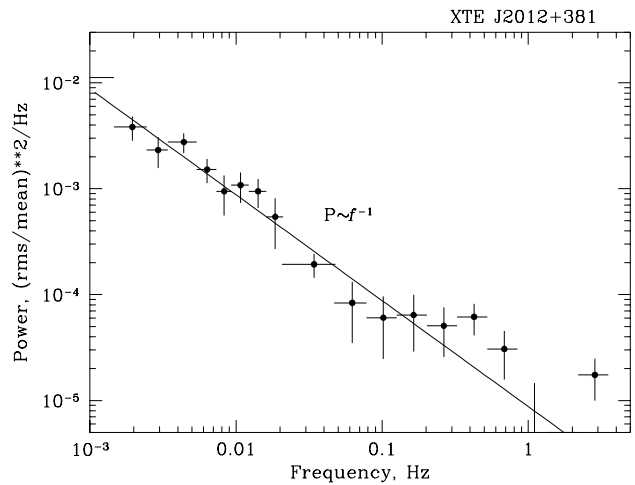}
\caption{Power spectrum of XTE~J2012+381 averaged over observations $\#2 
- \#17$ (PCA data, $3 - 20$ keV energy range). \label{power}}
\end{figure}

\begin{figure}
\epsfxsize=8cm
\epsffile[40 197 566 454]{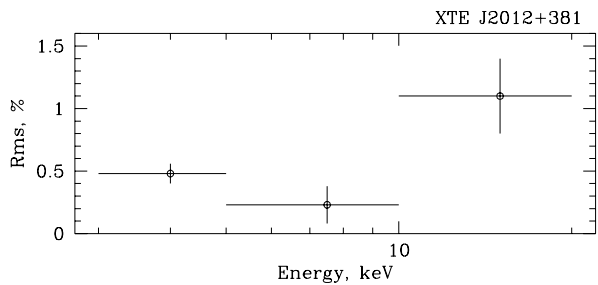}
\caption{The dependence of fractional rms variability of X-ray flux ($0.001 - 
1$ Hz) of XTE~J2012+381 on the photon energy. \label{rms}}
\end{figure}

The energy spectra of the X-ray transient were quite typical -- a dominant 
soft thermal component with a rather weak power law tail. Unfortunately, the 
transient was not bright enough for high energy instrument aboard RXTE 
(HEXTE), so we could not trace the energy spectrum of XTE~J2012+381 to 
energies higher than $\sim$80 keV. 

XTE~J2012+381 one more time demonstrated that the soft spectral component in
the spectra of X-ray Novae rise not instantly, but slightly after the power
law component. The similar behavior was observed in Nova Muscae 1991 
(\cite{muscae}), KS~1730-312 (\cite{ks1730}), GRS~1739-278 (\cite{1739}) and 
XTE~J1748-288 (\cite{1748}). The sequence of typical spectra is presented in
Fig. \ref{3spectra}. One can see that during the first observation the soft
component is not very strong (May 27, 1998), the second spectrum already
has much more dominant soft component (June 1, 1998) and the third spectrum
shows the weakening of the power law component (July 7, 1998). Broad band
spectrum, averaged over observations $\#2 - \#17$ is presented in Fig. 
\ref{broad}.

As it is clearly seen from Fig. \ref{power}, the average power density 
spectrum of the source is dominated by Very Low Frequency noise (VLFN) 
component, reasonably approximated by a simple power law model
($P\propto f^{\alpha}$) with a 
slope $\alpha=-1.08\pm0.06$ ($0.001 - 0.5$ Hz frequency range). We have 
not detected any statistically significant source flux variability at 
the frequencies higher than $\sim$ 1 Hz. The 2 $\sigma$ upper limit on 
the possible Lorentzian component at $f \sim 500 - 1000$ Hz is approximately 
2\% for $Q = 10$ and 1\% for $Q = 1$. In Fig. \ref{rms} the fractional rms 
amplitude of variability is shown as a function of the photon energy. In spite 
of a poor statistics (very low amplitude of X-ray variability), one can see
that there is an indication on the rise of the rms with energy. Such
dependence is  
quite typical for the emission of black hole candidates in the Very High 
State (VHS)(see discussion of such behavior in \cite{chur00}). 

\section{Discussion}

The the properties of the outburst of XTE~J2012+381 are very similar to 
what of the X-ray Novae. It has a FRED (fast-rise-exponential-decay) light 
curve, with secondary maxima, as typical for X-ray transients (\cite{chen97}).

The spectrum of the source can be well described by standard two-component 
model which is typical for black hole candidate X-ray binaries (\cite{tl95}). 
The evolution of the energy spectrum of XTE~J2012+381 can be characterized as 
a gradual increase of the contribution of the soft thermal component as the 
overall X-ray flux decreases.  

Our calculations show that the value of model parameter $R_{in} \cos i$, 
inferred from the soft spectral component approximation is nearly constant 
during the decay phase of the outburst (Table \ref{spec_pars}) in general 
agreement with earlier results of \cite{Zh98}. However we believe that it 
is premature to assume a constant radius of inner accretion disk during the 
outburst. \footnote{Note that no corrections to the electron scattering and 
general relativity were made in the soft spectral component model (\cite{ss73,
sht95}). This model assumes incorrect radial dependence of the disk effective 
temperature within $\sim 5$ gravitational radii. Thus the inferred value 
of the effective radius ${\rm R_{in}}$ should not be treated as the actual 
size of the optically thick emitting region.}

The PDS of XTE~J2012+381 is also very typical for the VHS of black hole 
candidates with dominant soft spectral component. The energy dependence of 
the fractional rms provided another hint at the universal correlation between 
the amplitude of the variability and relative strength of the hard spectral 
component.

\begin{acknowledgements}
This research has made use of data obtained 
through the High Energy Astrophysics Science Archive Research Center
Online Service, provided by the NASA/Goddard Space Flight Center.
\end{acknowledgements}

\end{document}